# Loss and Bandwidth Studies on Multimode Polymer Waveguide Components for On-Board High-Speed Optical Interconnects

Jian Chen, *Student Member, IEEE*, Nikolaos Bamiedakis, Peter P. Vasil'ev, Richard V. Penty, *Senior Member, IEEE*, and Ian H. White, *Fellow, IEEE*

*Abstract*—Optical interconnects play a key role in the implementation of high-speed short-reach communication links within high-performance electronic systems. Multimode polymer waveguides in particular are strong candidates for use in passive optical backplanes as they can be cost-effectively integrated onto standard PCBs. Various optical backplanes using this technology and featuring a large number of multimode polymer waveguide components have been recently demonstrated. The optimisation of the loss performance of these complex waveguide layouts becomes particularly important at high data rates (≥ 25 Gb/s) due to the associated stringent power budget requirements. Moreover, launch conditions have to be carefully considered in such systems due to the highly-multimoded nature of this waveguide technology. In this paper therefore, we present thorough loss and bandwidth studies on siloxane-based multimode waveguides and waveguide components (i.e. bends and crossings) that enable the implementation of passive optical backplanes. The performance of these components is experimentally investigated under different launch conditions for different waveguide profiles that can be readily achieved through fabrication. Useful design rules on the use of waveguide bends and crossings are derived for each waveguide type. It is shown that the choice of waveguide parameters depends on the particular waveguide layout, assumed launch conditions and desired link bandwidth. As an application of these studies, the obtained results are employed to optimise the loss performance of a 10-card shuffle router and enable ≥40 Gb/s data transmission.

*Index Terms*—optical interconnections, optical backplanes, polymer waveguides, multimode waveguides, refractive index, waveguide dispersion.

## I. INTRODUCTION

The continuing growth in data usage and storage driven by high-speed internet, cloud computing and "Big Data" and "Internet of Things" environments requires interconnections with higher bandwidth, lower power consumption and higher density for future data centres and high-performance computing (HPC) systems [1]–[3]. Optical technologies have a key role to play in this development enabling higher-speed and lower-latency interconnections at all communication levels from on-chip to rack-to-rack links [4]. Optical backplanes have attracted particular interest in recent years as they are considered to be the next level where optics replace conventional copper-based interconnects in real-world systems [5]. Significant research has been carried out in this area in the last decade, with various optical backplane systems demonstrated based on different optical waveguide technologies: fibre-optic [6], [7], planar glass waveguides [8], [9] and polymer waveguides [2]. In particular, multimode polymer waveguides are a promising candidate for use in board-level interconnects as they can be directly integrated on conventional printed circuit boards (PCBs) owing to the favourable material properties, and enable cost-effective system assembly with relaxed alignment tolerances owing to the relative large waveguide dimensions (typically 30 to 70 µm in width) [10], [11].

These optical backplanes feature a large number of on-board polymer waveguides and waveguide components to implement different passive interconnection architectures and enable complex layout topologies. Examples of such architectures include non-blocking shuffle routers [12] and multi-channel optical bus configurations [13]. Passive waveguide components such as crossings, bends and splitters/combiners are key elements in such backplanes in order to implement the complex layouts. For example, the 10-card shuffle router presented in [12] and illustrated in Fig. 1, includes ~1800 waveguide crossings and 100 90° bends on a 10×10 cm² substrate.

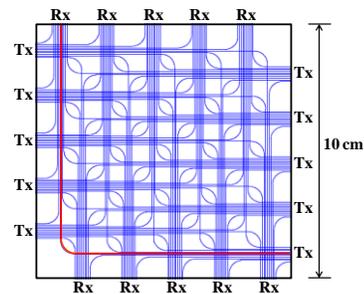

Fig. 1 Schematic of backplane illustrating the waveguide layout of a 10 card interconnection mesh (longest path indicated in red).

However, the optimisation of a particular waveguide layout is a challenging task due to the large number of waveguide components and their different behaviour with respect to the fundamental waveguide design parameters: waveguide dimensions and refractive index (RI) difference Δn. For example, the two basic components, waveguide bends and crossings, exhibit opposite behaviour with respect to the

J. Chen, N. Bamiedakis, Peter P. Vasil'ev, R. V. Penty and I. H. White, are with the Electrical Engineering Division, Engineering Department, University of Cambridge, CB3 0FA, Cambridge, UK (tel: 0044-1223748363; e-mail: jc791@cam.ac.uk).

waveguide RI difference Δn. Waveguide bends benefit from strong optical confinement at higher Δn [14], whereas waveguide crossings exhibit lower loss for lower RI difference Δn due to the smaller beam divergence at the waveguide intersections [15], [16]. Moreover, the multimode nature of this waveguide technology results in a different loss performance in these components which depends on the launch condition at their input. As a result, the optimum waveguide parameters for a particular layout depend on: i) the specific topology and number of components and ii) the type of input to be used. Given the continuous improvement in transmission data rates (≥25 Gb/s) in short-reach optical interconnects (current record of 40 Gb/s data transmission over 1 m long spiral waveguide [17]), the optimisation of the loss performance of complex waveguide layouts becomes particularly important due to the associated stringent power budget requirements. As a result, the optical losses need to be minimised while ensuring adequate bandwidth for the target data rate. In this work therefore, we present thorough loss and bandwidth studies on the basic waveguide components with different RI profiles that can be readily achieved through fabrication, and highlight the associated design trade-offs. The obtained results are used to optimise the loss performance of the shuffle router reported in [12], while ensuring adequate bandwidth for ≥40 Gb/s data transmission.

Although a number of studies on the loss performance of different waveguide components have been reported in recent years [18], [19], these focus only on the optimisation of the loss performance of a particular component (e.g. low loss in multimode waveguide crossings). Here, both bends and crossings are considered, while additionally their bandwidth performance under different launch conditions is reported for the first time. It is shown that the mode filtering properties of these components can provide bandwidth enhancement over the respective plain waveguides. The studies are carried out on siloxane-based waveguides which have been demonstrated in various prototype optical backplanes and are currently being used by a number academic and industrial research groups for their technology development [4], [20]. Moreover, the waveguide profiles studied can be readily achieved through fabrication, providing therefore practical results that highlight the associated design trade-offs and provide useful guidelines for system designers. The remainder of paper is structured as follows. Section II introduces the multimode polymer waveguides and waveguide components employed in this work, while section III presents the characterisation of their loss and bandwidth performance under different launch conditions. The obtained results are employed to optimise the loss performance of the 10-card shuffle router in section IV. Finally, section V draws the conclusions.

## II. MULTIMODE POLYMER WAVEGUIDE COMPONENTS

The waveguide samples employed in this work are fabricated from siloxane materials by Dow Corning (core: Dow Corning WG-1020 Optical Waveguide Core and cladding: XX-1023 Optical Waveguide Clad) using standard photolithographic processes [21]. The materials have been shown to exhibit very low intrinsic attenuation at datacommunication wavelengths (~0.03 dB/cm at 850 nm) and withstand solder reflow and environmental stability tests, exhibiting remarkable resistivity up to 350°C [22]. For this work, three different waveguide samples (denoted WG01, WG02 and WG03) are fabricated with slightly different RI profiles and dimensions (Fig. 2). The profile of WG01 and WG03 [Fig. 2(a) and Fig. 2(c)] correspond to the typical shape obtained through standard fabrication process, with a region of higher RI towards the top part of the waveguide. The profile of WG02 [Fig. 2(b)] is achieved by introducing an additional step in the fabrication process which enables diffusion of cladding monomers into the waveguide core. As a result, a smaller index difference Δn is obtained and a modified profile with the higher index region towards the waveguide bottom is achieved. Detailed studies carried out by Dow Corning have determined that these types of graded-index (GI) profiles can be reliably generated with low variability when many guides are formed. Fig. 2(d) summarises the basic waveguide parameters for each sample.

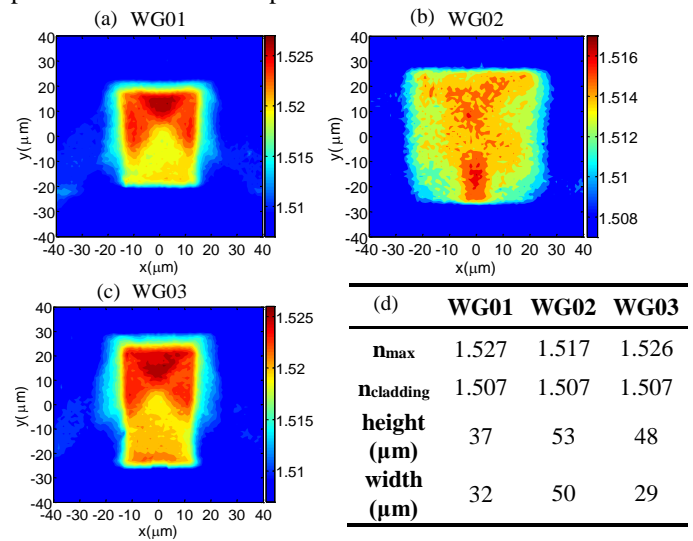

Fig. 2 (a-c) Measured RI profile of the 3 waveguide samples at 850 nm, and (d) Summary of the characteristics of 3 waveguide samples at 850 nm..

Each waveguide sample is fabricated on an 8-inch silicon substrate and it includes a number of test waveguide components:

- Waveguides with four 90° bends, two of which have constant radius (17 mm), while the other two have a varying radius of curvature R of 5, 6, 8, 11, 15 and 20 mm [Fig. 3(a)];
- Waveguides with a variable number of 90° and 45° crossings with a number of crossings of 1, 5, 10, 20, 40 and 80 [Fig. 3(b)];
- 16.25 cm long reference waveguides which include two 90° bends and one relatively long 180° bend [Fig. 3(c)];

The references waveguides have a similar shape and length as the waveguide components studied and are used to provide a reference loss and bandwidth performance against which the performance of the waveguide bends and crossings are compared. The waveguide facets are exposed using a dicing saw while no polishing steps are undertaken to improve their quality.

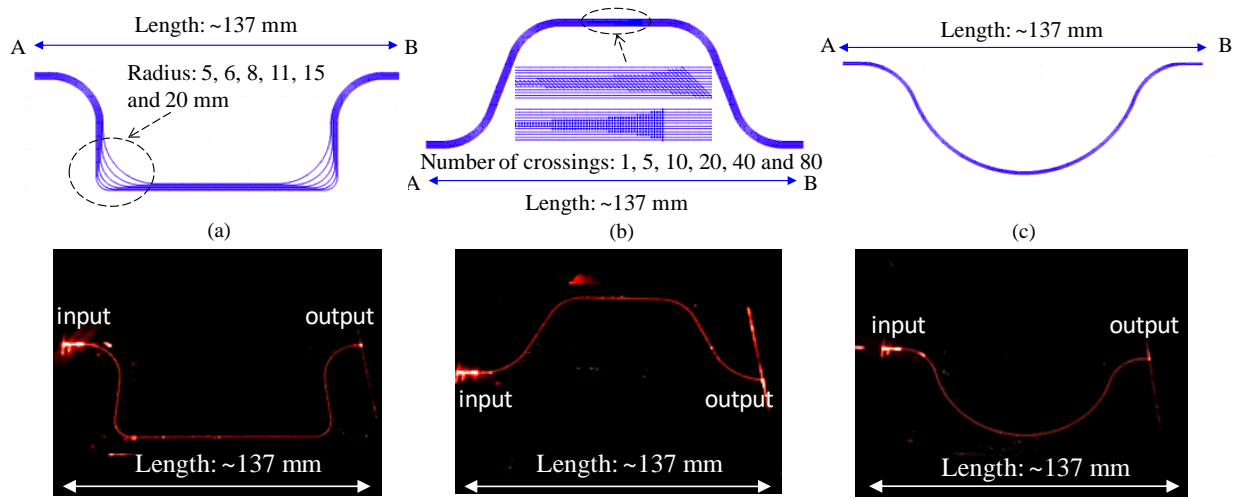

Fig. 3 Schematic of the waveguide components studied: (a) waveguide bends, (b) waveguide crossings and (c) reference waveguides, and respective photographs when illuminated with red light.

### III. Experimental Results

#### A. Power Transmission Studies

Loss measurements are conducted on the aforementioned multimode polymer waveguide components under different launch conditions. The employed launch conditions vary from a restricted launch (9/125 SMF) to a relatively overfilled launch (100/140 µm MMF with a mode mixer), providing a clear image of their loss performance. Three types of fibres are employed at the waveguide input to generate these different launch conditions: (i) a 9/125 µm single-mode fibre (SMF) input, (ii) a 50/125 µm GI multimode fibre (MMF) input [fibre numerical aperture (NA) = 0.2], and (iii) a 100/140 µm GI MMF input (NA = 0.29) used with a mode mixer (MM: Newport FM-1). The SMF launch excites the smallest number of modes, while the 100/140 µm MMF with the use of the MM results in the most overfilled launch. The 50/125 µm MMF input consist of a "medium" launch condition exciting a larger number of modes at the waveguide input than the 9/125 µm SMF but a smaller one than the 100/140 µm MMF launch, and corresponds to a more typical launch condition that could be encountered in a real-world system. The basic experimental setup for the loss measurements is shown in Fig. 4. For all measurements, a multimode 850 nm VCSEL is used as the light source, while a pair of microscope objectives is employed to couple the light into the appropriate fibre patchcord. Each launch condition is characterised prior to the waveguide loss measurements with near- and far-field measurements. Fig. 5 depicts the far-field intensities and the corresponding near-field images of the cleaved end of the three different input fibres. The -13 dB intensity points (5% value) in the far field plots indicate the NA of each employed input and this is found to be: 0.13, 0.18 and 0.26 for the SMF, 50/125 µm MMF and 100/140 µm MMF input respectively. It should be noted that the 9 µm SMF is not strictly single-moded at 850 nm and supports a small number of lower-order modes resulting in the far-field profile observed in Fig. 5(a).

The cleaved end of the input fibre is butt-coupled with the input facet of the waveguide component under test, while a 16× microscope objective (NA = 0.32) is used to collect the light at the waveguide output and focus it onto the head of a large area optical power meter (HP 81525A). The 16× lens is chosen as its NA is larger than that of the waveguide, preventing any mode selective loss at the waveguide output. For each measurement, the position of the input fibre is adjusted using a precision translation stage to maximise the power transmission through the waveguide component under test.

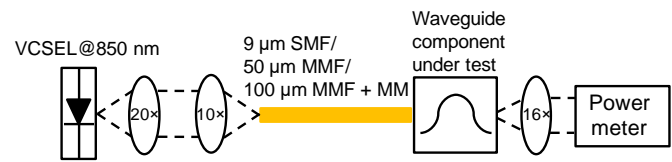

Fig. 4 Experimental setup for the power transmission measurements.

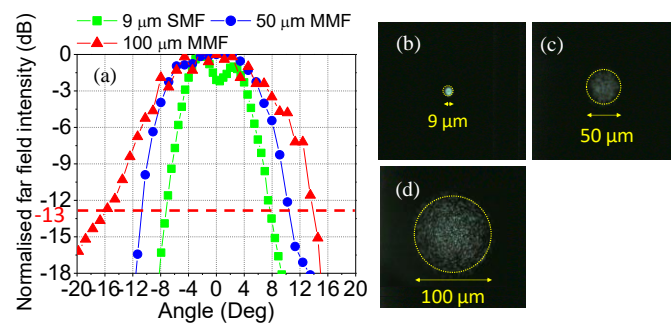

Fig. 5 (a) Far-field intensity of the 3 different launch conditions used in the measurements and respective near-field images of the fibre end: (b) a 9/125 µm SMF, (c) 50/125 µm MMF and (d) 100/140 µm MMF.

*I. Reference waveguides*

Fig. 6 shows the total insertion loss of the reference waveguides for the 3 waveguide samples under the different launch conditions employed. The total insertion loss values shown are the average values obtained from 6 parallel waveguides for each sample and include the coupling and propagation loss of the waveguides. The standard deviation in the measurements is found to be less than 0.1 dB for all inputs.

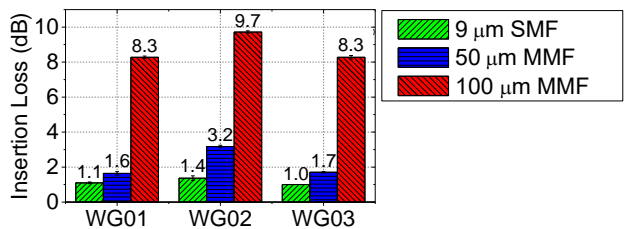

Fig. 6 Average insertion loss of the reference waveguides under the 3 launch conditions.

The obtained insertion loss values are used to calibrate the loss performance of the waveguide bends and crossings, as the different samples exhibit slightly different coupling loss and propagation loss values due to their different profiles. As a result, all loss values presented in the following sections for the waveguide components (bends and crossings) are normalised with respect to the insertion loss of the corresponding reference waveguide under the same launch condition. Therefore, the calculated loss values for the components presented below do not include coupling and propagation losses and indicate only their additional (excess) loss.

*II. Bent waveguides*

Fig. 7 shows the excess bending loss of the waveguides with the double 90° bends for the 3 samples as a function of the radius of curvature for the three launch conditions studied. As expected, the SMF input results in the lowest bending loss, whereas the 100 μm MMF input results in the highest value. The SMF mainly excites the lower-order modes at the waveguide input, while the 100 μm MMF couples the larger percentage of power to the higher-order modes which are more susceptible to radiation loss along the bends. Comparing the loss performance of the three samples, the results are in agreement with the expected behaviour when considering the waveguide parameters (size and RI difference) and the resulting light confinement in the plane of the bend. WG02 exhibits the largest bending loss due its smaller Δn value and larger waveguide width, while WG01 and WG03 exhibit roughly similar behaviour. WG03 has a slightly improved performance (by ~0.5 dB) due to the smaller core size in the plane of the bend (29 μm in comparison to 32 μm for WG01).

Table 1 summarises the minimum bend radius required to ensure bending losses < 1 dB for the 3 waveguide samples under the different launch conditions. The obtained values can be used as a design rule when drawing on-board waveguide layouts.

Table 1 Required radius for bending loss < 1 dB for the 3 samples under the different launch conditions studied.

| Sample | 9 μm SMF | 50 μm MMF | 100 μm MMF |
|---|---|---|---|
| **WG01** | > 6 mm | > 6 mm | > 8 mm |
| **WG02** | > 10 mm | > 11 mm | > 15 mm |
| **WG03** | > 5 mm | > 6 mm | > 8 mm |

A restricted launch requires a bend radius of at least 6 mm for WG01 and 5 mm for W03, while a more overfilled input requires a radius > 8 mm to ensure loss < 1 dB. For WG02, the respective values are 10 mm (restricted) and 15 mm (overfilled) indicating a larger space requirement.

*III. Waveguide crossings*

Fig. 8 shows the excess loss of the waveguides with the 90° and 45° crossings under the three launch conditions studied. Similar behaviour is observed for both types of waveguide crossings, with the losses for the 45° crossings being considerably higher as light is more likely to leak at the waveguide intersection for smaller crossing angles [15].

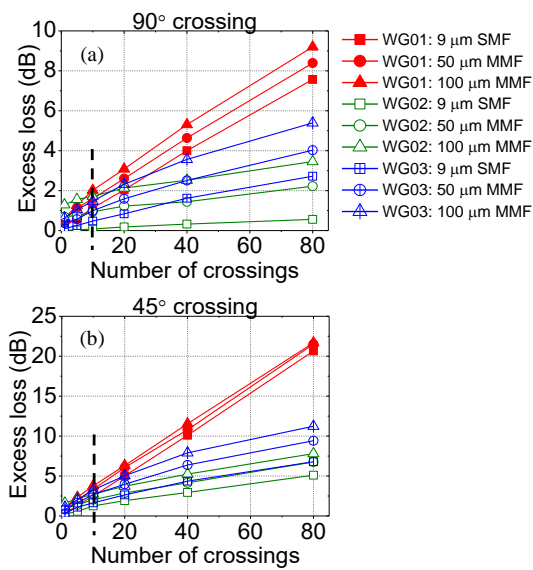

Fig. 8 Excess loss of the (a) 90° and (b) 45° crossings as a function of the number of crossings for the 3 samples and under the different inputs.

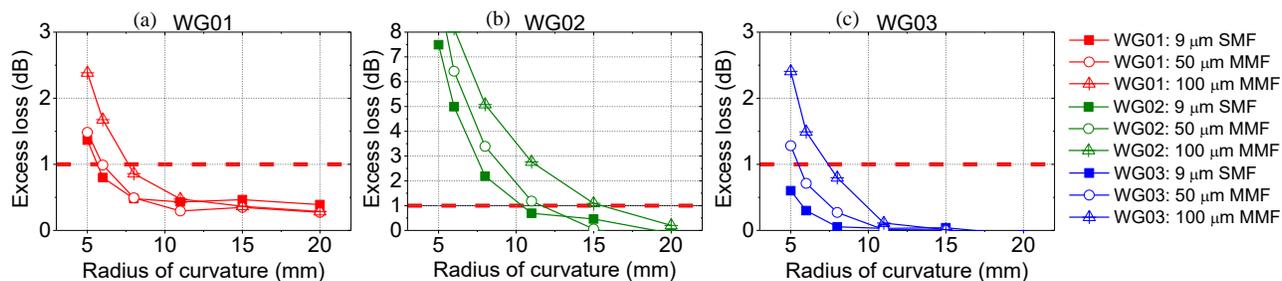

Fig. 7 Excess loss of the bent waveguides as a function of the radius of curvature for the 3 launches. The 1 dB insertion loss line is shown in red.

The 100 μm MMF launch results in the highest loss as it provides a more "overfilled" launch at the waveguide input, coupling more power into the higher-order modes which are more susceptible to crossing losses. As observed before, the crossing loss does not depend linearly on the number of crossings, as higher-order modes exhibit higher crossing loss and are therefore primarily attenuated at the initial crossings closer to the waveguide input [23]. Lower-order modes exhibit a lower attenuation coefficient and are left to propagate in the remaining crossings. As a result, two different slope approximations (loss per crossing) are extracted for each sample: for 1-10 crossings (slope: $k_1$) and for 20-80 crossings (slope: $k_2$) (Table 2) and can be used to estimate the total crossing loss for a particular optical path, featuring $x$ number of crossings:

$$\text{Excess loss} = \begin{cases} k_1 \cdot x & \text{if } x \leq 10 \\ k_1 \cdot 10 + k_2 \cdot (x-10) & \text{if } x > 10 \end{cases}$$

In addition, WG01 and WG02 exhibit the highest and lowest loss per crossing respectively due to their largest and smallest RI difference respectively.

Table 2. Loss per crossing approximation for the 90° and 45° crossings for 3 waveguide samples for the 3 launch conditions.

| | Input | 9 μm SMF | | 50 μm MMF | | 100 μm MMF | |
|---|---|---|---|---|---|---|---|
| | | Loss approximation in dB/crossing | | | | | |
| | Sample | $k_1$ 1-10 | $k_2$ 20-80 | $k_1$ 1-10 | $k_2$ 20-80 | $k_1$ 1-10 | $k_2$ 20-80 |
| 90° | WG01 | 0.098 | 0.092 | 0.122 | 0.096 | 0.155 | 0.101 |
| 90° | WG02 | 0.008 | 0.006 | 0.027 | 0.017 | 0.046 | 0.022 |
| 90° | WG03 | 0.042 | 0.031 | 0.070 | 0.040 | 0.092 | 0.050 |
| 45° | WG01 | 0.243 | 0.261 | 0.292 | 0.259 | 0.296 | 0.256 |
| 45° | WG02 | 0.114 | 0.053 | 0.119 | 0.065 | 0.125 | 0.067 |
| 45° | WG03 | 0.143 | 0.068 | 0.210 | 0.089 | 0.239 | 0.100 |

### B. Dispersion Studies

The bandwidth of the waveguide components under test is also measured for the different launch conditions using time-domain measurements. Fig. 9 illustrates the experimental setup used. A femtosecond erbium-doped fibre laser source (TOPTICA FemtoFiber Scientific) operating at 1574 nm and a frequency-doubling crystal (MSHG1550-0.5-1) are used to generate short pulses at the wavelength of 787 nm [full width at half-maximum (FWHM) of ~300 fs]. The optical pulses are coupled into the waveguide components via a 10× microscope objective (NA=0.25), a 50/125 μm MMF or a 100/140 μm MMF with a mode mixer, which resemble the types of launch conditions used in the power transmission studies. The 10× microscope objective launch provides a restricted launch condition with a Gaussian input spot with a FWHM of 5±1 μm. At the waveguide output a 16× microscope objective (NA = 0.32) is used to collect the light and deliver it to a matching autocorrelator [24], [25]. The received optical pulses after transmission over the link with and without the waveguide component are recorded and are used to estimate the waveguide 3 dB bandwidth by de-convolving the two frequency responses. A description of the methodology used is presented in greater detail in [24].

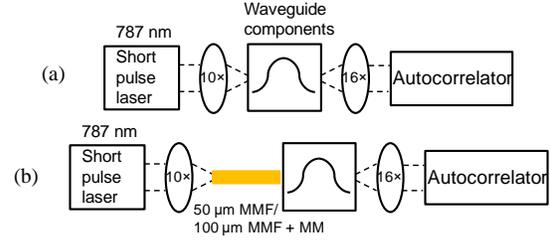

Fig. 9 Experimental setup for time-domain measurements for the waveguide link using (a) a 10× microscope objective input and (b) a 50/125 μm or a 100/140 μm MMF + MM launch.

Fig. 10 shows the bandwidth-length product (BLP) of the 3 reference waveguides under the different launch conditions As expected, the 10× lens launch results in the highest bandwidth (of >100 GHz×m), whereas the 100 MMF input yields the lowest BLP value due to the excitation of higher-order modes at the waveguide input. Moreover, WG02 exhibits the largest BLP value (~2.5× larger than WG01 and WG03) owing to its much smaller RI difference Δn. The WG01 and WG03 waveguide samples exhibit similar bandwidth performance with BLP values of ~45 GHz×m as their size and RI difference are not significantly different.

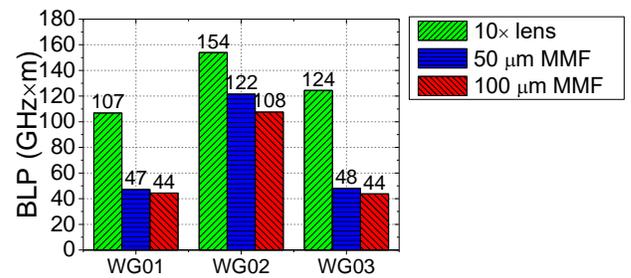

Fig. 10 BLPs of the 3 reference waveguides for the 3 launch conditions.

Far-field measurements are carried out on the reference waveguides to confirm the observations obtained from the bandwidth measurements. Fig. 11 shows the far-field intensity of the 3 reference waveguides under different launch conditions. WG02 has the narrowest far-field intensity profile for all inputs, while the WG01 and WG03 exhibit similar width in far-field intensity, confirming the bandwidth measurement results in Fig. 10.

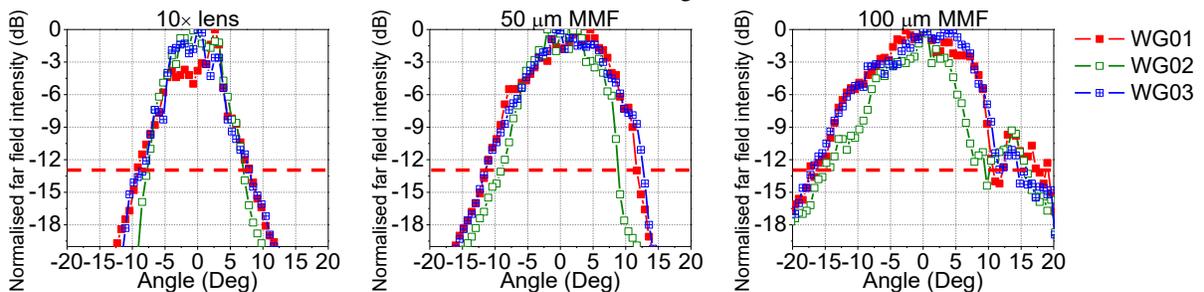

Fig. 11 Far field intensity of the 3 reference waveguides in the horizontal direction under (a) a 10× microscope objective, (b) a 50/125 μm MMF and (c) a 100/140 μm MMF launch.

Similar bandwidth measurements are carried out on the waveguide bends and crossings. Fig. 12 shows the obtained BLP values of the two components for the 3 waveguide samples under the 50/125 μm MMF launch. For all waveguide components, the bandwidth values increase for smaller radius of curvature and larger number of crossings, due to the stronger attenuation of the higher-order modes in the components. As a result, such components can also be used as mode filters, improving the BLP of a particular optical path, at the expense however of an increased insertion loss.

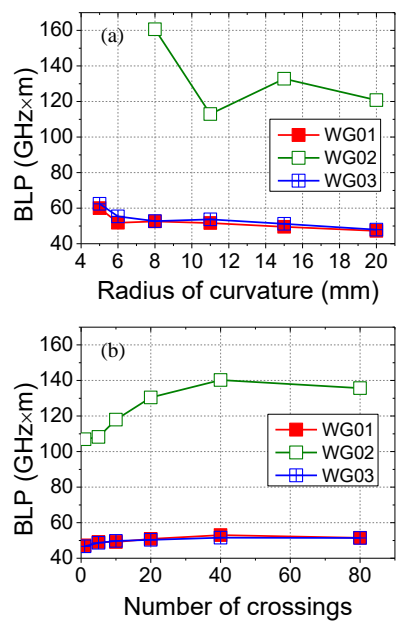

Fig. 12 BLPs of (a) the bent waveguides as a function of radius of curvature and (b) the waveguides with the 90° crossings as a function of the number of crossings for the 3 waveguide samples under the 50/125 μm MMF launch.

## IV. APPLICATION OF COMPONENT STUDIES

The obtained results highlight the design trade-offs associated with these elementary waveguide components. Samples WG02 exhibit the highest insertion loss and bending loss, but the lower crossing loss and larger BLP value. WG01 and WG03 samples exhibit lower insertion loss (mainly due to lower coupling loss) and crossing loss but higher bending loss and lower BLP. Roughly similar loss and bandwidth performance is recorded for these two samples (WG01 and WG03), indicating that the waveguide height has a relatively small effect on the performance of the components under test.

The obtained values are used to optimise the loss performance of the 10-card shuffle router shown in Fig. 1. To ensure reliable operation of the backplane, the worst-case optical path (indicated in red in Fig. 1) is considered. This includes 1 90° bend and 90 90° crossings. The backplane is expected to be interfaced with 50 μm MMF ribbons or directly butt-coupled with VCSEL arrays. As a result, the loss values obtained under the 50 μm MMF launch are used to calculate the loss performance expected for the 3 waveguide types. Assuming that enough space is available, the best loss performance for the worst-case optical path is calculated to be ~6 dB when the WG02 parameters are employed. In this case, a minimum radius of 12 mm is required for the 90° bends. If however, more stringent backplane size requirements are imposed, WG03 provides a better loss performance with an estimated total insertion loss for the worst-case path of ~6.1 dB. The corresponding minimum bending radius required is 8 mm in this case, resulting in a ~30 % reduction in size for each backplane side and ~52% in required area. Moreover, the use of waveguide parameters WG03 can offer a 2 dB loss improvement over the earlier backplane version reported in [12]. Finally, it should be noted that the bandwidth for the worst-case path is expected to be > 45 GHz×m even under a relatively overfilled launch, ensuring therefore adequate bandwidth for ≥ 40 Gb/s data transmission.

## V. CONCLUSION

Waveguide bends and crossings are fundamental components in the design of any complex on-board interconnection architecture employing multimode polymer waveguides. The optimisation of the loss performance of the on-board optical paths is a challenging task due to the large number of components used and their differing behaviour with respect to the fundamental waveguide design parameters (size, RI difference). Moreover, the loss behaviour of the waveguide components is strongly dependent on the launch conditions due to their highly-multimoded nature. Herein, thorough loss and bandwidth studies on the multimode polymer waveguides with different RI profiles and dimensions are presented. The studied parameters are typical waveguide profiles that can be obtained in siloxane-based waveguide components. The studies provide useful design rules for the insertion, bending and crossing loss of these components and highlight the underlying design trade-offs. It is shown that the optimisation of the loss performance of an optical path depends on the path topology, number of components and launch conditions used. As an example, the obtained results are used to optimise the loss performance of a 10-card shuffle router while ensuring adequate bandwidth for ≥ 40 Gb/s data transmission.

### ACKNOWLEDGMENT

The authors would like to acknowledge Dow Corning for providing the waveguide samples and EPSRC via the Complex Photonic Systems II (COPOS II) project for supporting the work. Additional data related to this publication is available at the University of Cambridge data repository.